\documentclass[aps,prl,twocolumn,showpacs,floatfix]{revtex4}
\usepackage{graphicx}
\begin{document}

\title{Pairing and Superfluid Properties of Dilute Fermion Gases at Unitarity}
\author{Vamsi K. \surname{Akkineni}}
\email{akkineni@uiuc.edu}
\affiliation{Department of Physics, University of Illinois at Urbana-Champaign,
         Urbana, IL 61801-3080\,}

\author{D. M. \surname{Ceperley}}
\affiliation{NCSA and Department of Physics, University of Illinois at Urbana-Champaign, Urbana, IL 61801-3080\,}

\author{Nandini \surname{Trivedi}}
\affiliation{Department of Physics, The Ohio State University, Columbus, OH 43210-1117\,}
\date{\today}
\
\begin{abstract}
We study the system of a dilute gas of fermions in 3-dimensions, with attractive interactions tuned to the unitarity point, using the non-perturbative Restricted Path Integral Monte Carlo (R-PIMC) method. The pairing and superfluid properties of this system are calculated at finite temperature. The total energy at very low temperature from our results agrees closely with that of previous ground-state Quantum Monte Carlo calculations. We identify the temperature $T^\ast\approx 0.70\epsilon_F$ below which pairing correlations develop, and estimate the critical temperature for the superfluid transition $T_c\approx 0.25\epsilon_F$ from a finite size scaling analysis of the superfluid density. 
\end{abstract}

\maketitle

The recent experiments, starting with generating a degenerate gas of cold atoms\cite{jin}, the use of Feshbach resonance to tune the effective interaction between the fermions\cite{jin2,ketterle}, the measurements of the gap to identify a pairing scale\cite{grimm} and the measurements of vortices\cite{vortices} to identify the superfluid state, have all given an impetus to the study of superfluidity in the BCS-BEC crossover regime. 

The nature of superconductivity or superfluidity in a many-particle system with an increasing pairing attraction was shown \cite{leggett,randeria-review} to interpolate smoothly between the BCS regime for weak attraction between the fermions, and the Bose-Einstein condensation (BEC) regime for strong coupling. In the weak coupling regime, the Cooper pair size is much larger than the interparticle spacing, and the simultaneous pairing and condensation of fermions is well described by the BCS theory. In the strong coupling regime, fermions form strongly bound bosonic molecules at a pairing temperature scale $T^\ast$ that is considerably higher than the BEC temperature $T_c$.

In a two species system of fermions with an attractive interaction between them, the unitary point is defined  by the divergence of the zero energy \textit{s}-channel scattering length, $a_s$, for two particles in free space. The unitary point is interesting since it describes a strongly correlated system with universal properties\cite{ho}. In general, there are two length scales that define the system: the interparticle spacing $\propto n^{-1/3}$, and the scattering length $a_s$ which contains information about the interaction potential. At the unitary point however, since $a_s$ diverges, all the properties of the system are described by a single length scale $k_F^{-1}$ or energy scale $\epsilon_F$ where $k_F=(3\pi^2\hbar^3n)^{1/3}$ is the Fermi momentum of the corresponding non-interacting system.

At the unitarity point there is no small parameter, therefore well controlled numerical methods are needed to calculate the properties of the system. Here we present the first calculation of the superfluid density and other pairing correlations as a function of temperature at the unitary point in a {\em continuum} model of a two component fermion gas with attractive pairwise interaction between the species. Our main results are: (i) an accurate determination of the temperature dependent internal energy which makes it a viable tool for thermometry for cold atoms; (ii) determination of the pairing scale $T^\ast\simeq 0.7\epsilon_F$ from growth of density correlations of opposite spin fermions; (iii) determination of the condensation scale $T_c\simeq 0.25 \epsilon_F$ from a finite size scaling analysis of the superfluid density. We use the restricted path intergal Monte Carlo (R-PIMC) technique, which is the fixed-node extension of the continuum PIMC method to fermionic systems\cite{pimc,rpimc}. This technique has recently been used to study helium-3, electron-hole liquids, and many body hydrogen\cite{rpimc-applications}.

We consider an unpolarized system of two spins (or two hyperfine species) of particle density $n$. For the bare two-particle interaction at unitarity, we use the potential (also used by \cite{gfmc},)
\begin{equation}
v(r)=-{2\hbar^2\over m}{\mu^2\over \textup{cosh}(\mu r)}\,,\label{modpot}
\end{equation}
where $2/\mu$ is the effective range of the potential. This potential has its only bound state at zero energy, with the eigenfunction given by ${\rm tanh}(\mu r)/r$, and a divergent scattering length $a_s$. We measure distances in units of $r_0$, the average interparticle spacing, which gives the density $n=3/(4\pi r_0^3)\simeq 0.238$ and $k_Fr_0=1.919$. The energy scale is $\epsilon_0=\hbar^2/mr_0^2$ and the Fermi energy $\epsilon_F=1.841\epsilon_0$. All the following results were obtained with $\mu r_0=12$. In the dilute limit, with the interparticle spacing much larger than the range of the potential $\mu r_0\gg 1$, the exact form of the potential is unimportant; only the scattering length matters.

\textit{Method:} For a system of distinguishable quantum particles, the density matrix in configuration space can be written as a path integral over coordinate space variables at discrete imaginary time intervals as,
\begin{equation}
\rho(R_0, R_{\rm M}; \beta)=\int dR_1\cdots dR_{{\rm
M}-1}\,\exp\biggl[-\sum_{m=1}^M S^m\biggr]\,.\label{pathint2}
\end{equation}
Here $R_m$ denotes a configuration of $N/2$ up (labeled with $\uparrow$) and $N/2$ down (labeled with $\downarrow$) fermions at the $m^{th}$ timeslice. The total extent in the imaginary time direction, $\beta=1/k_B T$, is divided into $M$ timeslices each of size $\tau$ so that $\beta=M\tau$.  The link action $S^m$ is the sum of the kinetic-action terms and potential-action $U^m$, given by,
\begin{equation}
S^m={3N\over 2}{\rm ln}(4\pi\epsilon_0\tau)+{(R_m-R_{m-1})^2\over
4\epsilon_0\tau}+U^m\,.\label{kinact}
\end{equation}
Eq.~\ref{pathint2} is the basis for a quantum-classical isomorphism between a system of quantum particles and a classical system of interacting polymers. Each particle path is interpreted as a polymer, with the beads of the polymer connected with springs described by the kinetic action. The potential-action describes the interaction between beads of different polymers.

For indistinguishable fermions, the density matrix is anti-symmetrized over particle permutations,
\begin{equation}
\rho_{F}(R,R';\beta)={1\over N!}\sum_\mathcal{P}(-1)^\mathcal{P}
\rho(\mathcal{P}R,R';\beta)\,.\label{aspathint}
\end{equation}
In the classical picture, a permutation $\mathcal{P}$ involving $n$ particles corresponds to the cutting and combining of the corresponding $n$ polymers into one large polymer. As paths extend in space at low temperatures these quantum exchanges become more likely. 

For fermions however, a straightforward evaluation of the fermion density matrix given by (\ref{aspathint}) leads to the fermion sign problem. It arises due to the cancellation, at low temperature, of approximately equal contributions from the positive and negative sign permutations, leading to an exponentially vanishing signal-to-noise ratio in the Monte Carlo calculation \cite{rpimc}. 

The density matrix is a function of $2dN+1$ variables (the end configurations and time,) and can be solved for in a region of `space-time' with restricted paths if the initial condition $\rho_F(R,R';0)={(-1)^\mathcal{P}\over N!}\sum_\mathcal{P} \delta(\mathcal{P}R-R')$, and Dirichlet boundary conditions for $\tau\ne0$, are fully specified \cite{rpimc}. It is natural to choose zero boundary conditions to define the path restriction region, so that knowledge of the nodal surfaces of the density matrix allows computation of the thermodynamic properties of the system. 

To enable the computation, we utilize a trial density matrix constructed to have reasonable physical and topological properties, that defines the nodal surfaces to restrict the paths. The R-PIMC method evades the sign problem for calculations which involve the diagonal elements of the fermion density matrix. Due to the positivity of the density matrix on the diagonal, $\rho_F(R,R;\tau)>0$, we can disregard any paths that start at a negative permutation of $R$ since such paths have to cross a node at least once to end at $R$. We sum over only the positive (even) permutations, and keep only the positive node avoiding paths thus overcoming the sign problem. The permutation sum as well as the integrals in (\ref{aspathint}) are evaluated with the Metropolis Monte Carlo method.

In the high temperature limit, it can be shown \cite{nodes} that the interacting density matrix is well approximated by the free particle density matrix, $\rho_F(R,R^\ast;\tau)= (4\pi\epsilon_0 \tau)^{-{dN/2}}{\bf det}\bigl[\exp\{{-(r_i-r_j^\ast)^2/4\epsilon_0 \tau}\}\bigr]$. In the low-temperature limit, the contribution of the excited states in the spectral expansion of the fermion density matrix, $\rho_F(R,R';\tau)=\sum_n \Psi_n(R)\Psi_n^\ast(R')\exp(-\tau E_n)$, are exponentially damped relative to the ground state. The density matrix can then be approximated for a non-degenerate ground state as,
$\rho_F(R,R';\tau\rightarrow \infty)=\Psi_0(R)\Psi_0^\ast(R')$.
For the ground state trial wave function, we use a BCS-like antisymmetrized product of pairing functions \cite{footnote2}, $\phi(r^\uparrow-r^\downarrow)$ between opposite spin fermions, and given by $\Psi_0(R)={\bf det}[\phi(r^\uparrow_i-r^\downarrow_j)]$.

\begin{figure}
\includegraphics*[scale=1.0,angle=0]{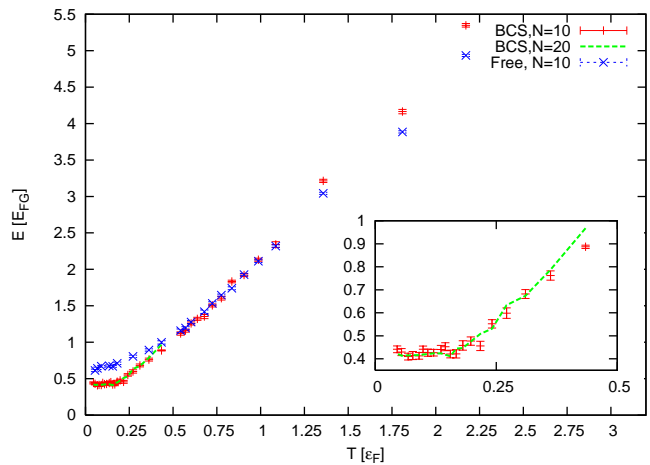}
\caption{\label{ENERGY} The energy per particle in units of $\epsilon_{FG}=(3/5)\epsilon_F$ as a function of $T$ for $N=10$ and $20$ (dashed line; error bars not shown but comparable to $N=10$) computed using the BCS-like trial density matrix (`BCS'). For comparison the energy obtained by using the non-interacting particle trial density matrix (`Free') is also shown. }
\end{figure}
\textit{Results:} The total energy vs. temperature is shown in Fig. \ref{ENERGY} for $N=10$ and $N=20$. The BCS-like trial density matrix, in addition to its enabling the physics of pairing and condensation, is further justified because it gives the lowest total energy at low temperature of all the trial density matrices that we have tested. We also find good agreement of our energy extrapolated to $T=0$ with the value of $E_o/N=0.44(1)\epsilon_{FG}=0.26\epsilon_F$ obtained from the fixed node Green function Monte Carlo method\cite{gfmc,gfmc2}, where $\epsilon_{FG}={3/5}\epsilon_F$ is the average energy per particle of the non-interacting Fermi gas. 

Estimation of $T^\ast$:
The first indication of the crossover pairing scale $T^\ast\approx 0.70\epsilon_F$ is obtained from a comparison of the energy obtained by using the free particle trial density matrix, and the BCS-like density matrix as shown in Fig. \ref{ENERGY}. For $T\ge T^\ast$ the free particle density matrix gives the lowest energy, whereas in the opposite regime $T\le T^\ast$ the BCS density matrix gives the lowest energy.

\begin{figure}
\includegraphics*[scale=1.0,angle=0]{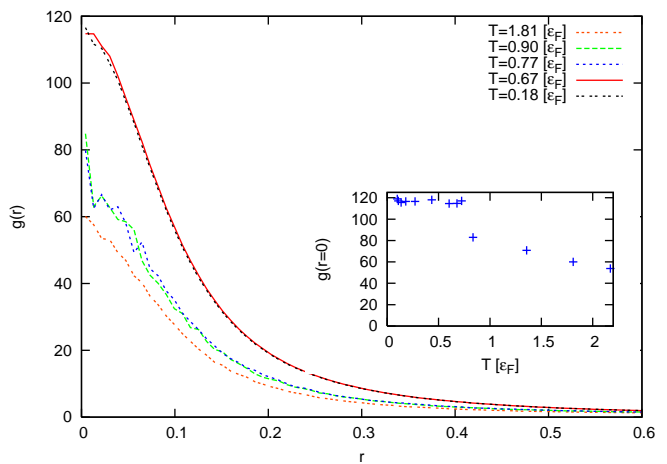}
\caption{\label{GRR} The pair correlation function of the two species of fermions at different values $T$. Note the sharp increase in pair correlation from $0.67T_F$ to $0.54T_F$.}
\end{figure}

Further evidence of the pairing scale comes from $g_{\uparrow-\downarrow}$ pairing correlations as seen in Fig. \ref{GRR}. As the temperature is lowered, there is a strong enhancement of the on-site density correlations for opposite spins at $T\sim 0.67 \epsilon_F$ (see inset) which identifies a pairing scale below which strong pairing correlations exist at temperatures well above the actual superfluid transition.

In previous work\cite{sademelo} a functional integral approach was used to estimate $T^\ast$ as a function of $1/(k_Fa_s)$ from a saddle point analysis of the gap and number equations. At unitarity they found $T^\ast\simeq 0.57 \epsilon_F$. Upon including fluctuations to quadratic order around the saddle point, the transition was suppressed, especially around unitarity and beyond, to a lower $T_c$. They estimated $T_c\simeq 0.22\epsilon_F$ at unitarity. The inclusion of the fluctuations around the saddle point to fourth order showed that the variations of $T_c$ is controlled by the coherence length $k_F\xi_0$. In the BCS regime $k_F\xi_0$ is exponentially large, while in the BEC regime, for a dilute Bose gas, $k_F\xi_0$ grows as a power law. In both these regimes, $T_c$ variations are small. However, around the unitary point, $k_F\xi_0\sim {\cal O}(1)$ and the fluctuations in $T_c/\epsilon_F$ are of order unity. This is precisely where simulations are of greatest value since the system is in a strongly correlated regime inaccessible to perturbation theory.
\begin{figure}
\includegraphics*[scale=1.0,angle=0]{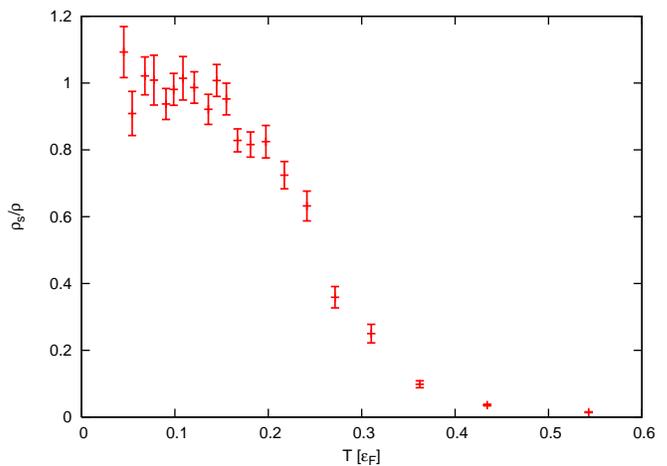}
\caption{\label{SFD} The superfluid fraction $\rho_s/\rho$ as a function of $T$ for $N=20$.}
\end{figure}
\begin{figure}
\includegraphics*[scale=1.0,angle=0]{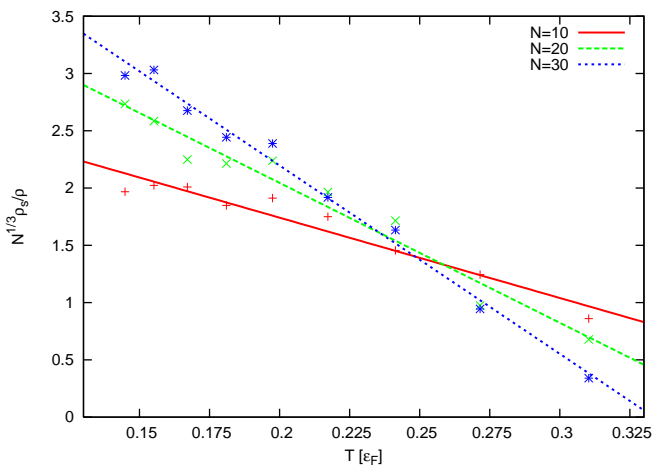}
\caption{\label{SFDSC} The scaled, linearized, superfluid density, $N^{1/3}\rho_s/\rho$,
as a function of $T$ for $N=10,20,\textrm{and } 30$. The intersection of the three lines gives the superfluid transition temperature as, $T_c\approx0.25T_F$.}
\end{figure}

We compute the superfluid density to directly estimate the critical temperature $T_c$ for the superfluid transition. The superfluid density $\rho_s$ for an $N$-particle system, shown Fig. \ref{SFD}, is calculated from the winding number estimator \cite{pollack},
\begin{equation}
{\rho_s\over \rho}={\langle{\bf W}^2\rangle\over 2\epsilon_0\beta N}\,.\label{windnum2}
\end{equation}
Here, $\bf{W}$ is the winding number defined as the number of times periodic boundary conditions are invoked as paths start from $R_i$ and end at a periodic image $R_{\mathcal{P}i}$. 

We start by reviewing the Josephson relation which is a hyperscaling relation between the correlation length exponent $\nu$ defined by $\xi\sim t^{-\nu}$ and the superfluid density exponent $\zeta$ defined by $\rho_s\sim t^\zeta$, where $t=(T-T_c)/ T_c$ is the reduced temperature. The singular part of the free energy density near the transition is $f_s(t)\sim \xi^{-d}$, and since $\rho_s(\nabla \phi)^2\sim f_s$, we obtain $\rho_s\sim \xi^{-(d-2)}\sim t^{\nu(d-2)}$ which directly gives the Josephson relation\cite{goldenfeld} $\zeta/\nu=d-2$. In a finite system, the scaling hypothesis\cite{goldenfeld} $f_s(t,L)=L^{-d}{\cal F}[L/\xi(t)]$ states that the free energy depends on $t$ only through the ratio of the system size $L$ and the bulk correlation length $\xi$. Using the Josephson relation, this implies that $\rho_s(t,L)=L^{2-d}Q[{L/\xi(t)}]$. In $d=3$, linearizing the function $Q$ near $t=0$ and using $L\sim N^{1/3}$ leads to
\begin{equation}
Q\biggl[{L\over\xi(t)}\biggr]=N^{1/3}{\rho_S(t)\over \rho}\approx Q(0)+qN^{1/3\nu}{T-T_c\over T_c}
\end{equation}
where $q$ is a constant. In Fig. \ref{SFDSC} we plot $N^{1/3}\rho_s(t)/\rho$ vs $T$ for several system sizes $N$. At the transition temperature $T_c$, the size dependence vanishes and all the curves meet at a point which determines the critical temperature $T_c\approx0.25\epsilon_F$. Our result agrees well with the estimate of $T_c\approx 0.22\,\epsilon_F$ \cite{sademelo} obtained by including fluctuations around the saddle point.

The transition temperature has also been calculated using lattice Monte Carlo techniques\cite{bulgac,burovski}. Our estimate is however higher than the lattice Monte Carlo estimate of Burovski {\it et al.} of $T_c\approx 0.15\;\epsilon_F$ \cite{burovski}. The main distinction of our path integral Monte Carlo is that we work directly in the continuum so the unitary limit is perfectly well-defined. The lattice simulations have to extrapolate $T_c$ to the zero filling factor limit in order to get the correct behavior in the unitary limit. On the other hand, for the attractive-U Hubbard model there is no sign problem\cite{negu-nt} which is a definite advantage. However, given the good agreement of our finite temperature R-PIMC method with the zero temperature GFMC results for the energy, we believe we have an accurate description of the nodal surface. The R-PIMC method can further be used to study the competition between superfluidity and magnetization in unequal fermion populations.

One of the difficulties in the experiments is to determine the temperature precisely. 
Usually a Maxwell Boltzmann fit to the excited atoms yields an estimate. However, at low temperatures when the number of atoms in the excited states is greatly reduced such an estimate becomes unreliable\cite{thomas}. The strong dependence of the energy on temperature above $T_c$ seen from our results, indicates that measurements of the mean field energy can be converted to a temperature scale, with appropriate corrections for a trap using a local density approximation.

We would like to thank Mohit Randeria, Roberto Diener and Jason Ho for valuable discussions. This work has been supported by the NSF under the Grant No. DMR-0404853. Computer time was provided by the Materials Computation Center at the University of Illinois at Urbana-Champaign.


\begin{thebibliography}{}
\bibitem{footnote2} For the pairing function  $\phi(r^\uparrow_i-r^\downarrow_j)$ we choose a gaussian with one free parameter that is optimized; the results are not sensitive to this specific choice.
\end{thebibliography}
\end{document}